\documentclass[aps,prc,twocolumn,showpacs]{revtex4}

\usepackage{graphicx}
\usepackage{dcolumn}

\begin{document}


\title{Electromagnetic Transition Form Factors of Light Mesons}


\author{Pieter Maris}
\email[]{pmaris@unity.ncsu.edu}
\affiliation{Department of Physics, 
North Carolina State University, Raleigh,  NC 27695-8202}

\author{Peter C. Tandy}
\email[]{tandy@cnr2.kent.edu}
\affiliation{Department of Physics, Center for Nuclear Research,
Kent State University, Kent, OH 44240}

\date{\today}

\begin{abstract}
We study selected meson transition processes and associated form
factors within a model of QCD based on the Dyson--Schwinger equations
truncated to ladder-rainbow level.  The infrared structure of the
ladder-rainbow kernel is described by two parameters; the ultraviolet
behavior is fixed by the one-loop renormalization group behavior of
QCD.  The work is restricted to the $u$ and $d$ quark sector and
allows a Poincar\'e-covariant study of the radiative decays:
\mbox{$\rho \to \pi \gamma$}, \mbox{$\omega \to \pi \gamma$}, and
\mbox{$\pi^0 \to \gamma \gamma$}.  Particular attention is paid to the
form factors for the associated transitions: \mbox{$\gamma^* \pi^0 \to
\gamma$}, \mbox{$\gamma^* \pi^0 \to \gamma^*$}, \mbox{$\gamma^* \pi
\to \rho$} and \mbox{$\gamma^* \rho \to \pi$}.  The latter two
processes are of interest as contributors to meson electroproduction
from hadronic targets away from the $s$-channel resonance region.  We
use the present QCD model to explore limitations to the assumption
that coupling to a $t$-channel virtual $\bar{q}q$ correlation can be
modeled as meson exchange.
\end{abstract}



\maketitle

\section{\label{sec:intro}
Introduction}
At a soft scale, the electromagnetic content of hadrons reveals the
distribution of current generated by the interacting constituent
dressed quarks.  This useful insight into non-perturbative hadron
dynamics is expressed in terms of electromagnetic form factors
obtained from exclusive processes such as elastic lepton scattering or
leptonic transitions to specific hadronic states.  At a hard scale,
typified by present deep inelastic scattering experiments, much of the
final state non-perturbative dynamics is reflected in the initial
distributions of quark and gluon partons which are probed
perturbatively.  The unraveling of this information from structure
functions will require a connection with non-perturbative QCD
calculations and models.  Although some lattice QCD studies have begun
to produce moments of structure
functions~\cite{Gockeler:1996wg,Gockeler:2001us}, the opportunities
are very limited at present.  Useful quantities for calibration of
models of QCD are the electromagnetic elastic and transition form
factors of hadrons.

The most extensive hadronic models were designed to study the mass
spectrum and decays and often contain elements that limit their use
for developing electroweak form factors.  Examples include:
non-relativistic kinematics, a lack of manifest Poincar\'e covariance,
no quark sea, no dynamical gluons, no QCD renormalization group
behavior for evolution of scale, and no confinement of quarks.
Covariant relativistic field theory models that are simpler than QCD,
but respect dynamical chiral symmetry breaking, have long been
utilized for the modeling of soft physics such as the hadronic mass
spectrum.  For example, such progress has been reviewed within the
Nambu--Jona-Lasinio model~\cite{Vogl:1991qt} and the Global Colour
Model~\cite{Cahill:1998rs,Tandy:1997qf}.  In the case of the latter, soft
electromagnetic form factors of mesons and meson transitions have been
considered~\cite{Tandy:1997qf}.  For hard electromagnetic processes,
studies of deep inelastic scattering within the Nambu--Jona-Lasinio
model~\cite{Weigel:1999pc,RuizArriola:2001rr} have helped clarify some
of the issues confronting work within a quark field theory format.

Here we summarize recent progress in the soft QCD modeling of light
meson electromagnetic transition form factors based on the set of
Dyson--Schwinger equations [DSEs] of the
theory~\cite{Roberts:2000aa,Alkofer:2000wg}.  The model we apply here
has been previously shown to give an efficient description of the
masses and electroweak decays of the light pseudoscalar and vector
mesons~\cite{Maris:1997tm,Maris:1999nt}.  Here we work in the $u$ and
$d$ quark sector and treat electromagnetic processes involving $\pi$,
$\rho$ and $\omega$.  This covariant approach accommodates quark
confinement and implements the QCD one-loop renormalization group
behavior.  The performance of this model for deep inelastic scattering
phenomena can be gauged from that of a simplified version that has
recently produced excellent results for the pion valence quark
distribution amplitude~\cite{Hecht:2000xa}.
 
An issue that arises in studies of electromagnetic interactions with
hadrons is vector meson dominance [VMD].  Within a quark-gluon model
that dynamically produces the vector meson pole in the dressed
photon-quark vertex, the validity and effectiveness of extrapolating
such a mechanism to nearby momentum domains can be tested.  The pion
charge radius and low-$Q^2$ charge form factor have recently been
explored from that perspective~\cite{Maris:1999bh}.  The empirical
effectiveness of the simple VMD assumption in that case is much
greater than its faithfulness to the underlying dynamics.  A similar
example is found here in the transition \mbox{$\gamma^* \pi^0 \to
\gamma$}.  The vector meson resonance pole term extrapolated to the
photon point produces an estimate of the $\rho$ radiative decay
coupling constant $g_{\rho \pi \gamma}$ in terms of the $\pi$ decay
coupling constant $g_{\pi \gamma \gamma}$ that is accurate to within a
few percent.  For a large range of spacelike momentum, we find the
shape of the form factor for this transition to be consistent with a
monopole with mass scale $m_\rho$; this is also consistent with
analyses of the asymptotic behavior.

When a process like the above involves a $\bar{q}q$ correlation with a 
spacelike invariant mass, there arises the question of how accurately 
it can be modeled by meson exchange when the meson is a bound state
defined at a unique (timelike) invariant mass.   This question is
here brought into sharper focus by examination  of form factors for 
transition processes $\gamma^* M M'$ that arise in meson exchange 
models~\cite{Sato:2000jf}.   Here $M$ is a virtual $t$-channel $\bar{q}q$ 
correlation such as might be provided by a hadronic target, and $M'$ is
a physical produced meson.  Our model of the underlying quark-gluon
dynamics is used to investigate the extent to which the virtuality of
$M$ influences the $\gamma^* M M'$ form factor that should be employed
in meson exchange models, given that the employed $t$-channel
propagator for $M$ is of the standard point meson type.  We consider
the cases where $M$ can be vector or pseudoscalar.   The domain of
accuracy of the meson exchange assumption is explored.

In Sec.~\ref{sec:DSEapproach} we use the generic coupling of a
pseudoscalar bosonic object to a pair of vector bosons in impulse
approximation to define our notation.  There we introduce the required
dressed quark propagator from the QCD quark DSE in rainbow
approximation, and the required vertex amplitudes from the
Bethe--Salpeter equation [BSE] in ladder approximation.  We also
outline there the phenomenological infrared content and
parameterization of the model and summarize the resulting pseudoscalar
and vector meson properties.  We specialize to the \mbox{$\pi \gamma
\gamma$} process in Sec.~\ref{sec:pigg}, discussing both the coupling
constant and two different transition form factors and their
asymptotic behavior.  In Sec.~\ref{sec:rhopig} the \mbox{$\rho \pi
\gamma$} coupling is discussed from three different aspects: the
coupling constants and widths for radiative decay of $\rho$ and
$\omega$, the form factor for the \mbox{$\gamma^* \pi \to \rho$}
transition, and the form factors for \mbox{$\gamma^* P \to \rho$} and
\mbox{$\gamma^* V \to \pi$} where $P$ and $V$ are virtual $\bar{q}q$
objects having the quantum numbers of ground state pseudoscalar and
vector mesons respectively.  Here the connection to $t$-channel meson
exchange is explored.  A summary can be found in
Sec.~\ref{sec:summary}.

\section{\label{sec:DSEapproach}
Pseudoscalar-Vector-Vector processes in impulse approximation}
Here we analyse pseudoscalar-vector-vector processes (PVV)
characterized by a generic PVV vertex that, in QCD in impulse
approximation, is described by the quark loop integral
\begin{eqnarray}
 \Lambda^{PVV}_{\mu\nu}(P;Q) &=& 
           N_c \int^\Lambda_q \!\! {\rm Tr}\big[
        S^a(q_a) \, \Gamma_{P}^{a\bar{b}}(q_a,q_b) \, S^b(q_b) 
\nonumber \\ &&  \times
        \Gamma^{b\bar{c}}_\mu(q_b,q_c)\, S^c(q_c) \, 
        \Gamma^{c\bar{a}}_\nu(q_c,q_a) \big] \;.
\label{eq:genPVV}
\end{eqnarray}
Here $\Gamma_{P}^{a\bar{b}}(q_a,q_b)$ is the pseudoscalar vertex
function for a pseudoscalar coupling to an outgoing quark with
momentum $q_a$ and flavor $a=u,d$, and an incoming quark with momentum
$q_b$ and flavor $b=u,d$; similar definitions hold for the two vector
vertex functions $\Gamma_\mu$ and $\Gamma_\nu$.  The external momenta
are $Q = q_c-q_a$ and $P = q_b-q_c$ for the vector vertices, and $K =
-(P+Q) = q_b-q_a$ for the pseudoscalar vertex, all flowing into the
triangle diagram.

The notation \mbox{$\int^\Lambda_q \equiv \int^\Lambda d^4
q/(2\pi)^4$} stands for a translationally invariant regularization of
the integral, with $\Lambda$ being the regularization mass-scale.  The
regularization can be removed at the end of all calculations, by
taking the limit \mbox{$\Lambda \to \infty$}.  It is understood that
the same regularization is applied to the calculated dressed quark
propagators $S(q)$ and vertex functions $\Gamma$ appearing in
Eq.~(\ref{eq:genPVV}), and that the quark propagators are renormalized
at a convenient spacelike momentum scale before taking the limit
\mbox{$\Lambda \to \infty$}.

This generic PVV vertex can be specialized in an obvious way to such
processes as the \mbox{$\pi^0 \to \gamma \gamma$} decay, radiative
decays such as \mbox{$\rho \to \pi \gamma$}, and transitions such as
\mbox{$\gamma^* \pi \to \rho$}.  In general, one has to add loop
integrals corresponding to the different orderings of the vertices and
the various flavor-labeled components of neutral mesons, in order to
get the vertex describing an actual physical process.

The form factor associated with the generic PVV vertex is identified
from the general form
\begin{eqnarray} 
 \Lambda^{PVV}_{\mu\nu}(P;Q) &=& C \; 
        F_{PVV}\left((P+Q)^2,P^2,Q^2\right) 
\nonumber \\ &&{} \times
        \epsilon_{\mu\nu\rho\sigma} \, P_\rho \, Q_\sigma
\label{eq:ffPVV}
\end{eqnarray}
normalized in such a way that $F_{PVV}=1$ for on-shell external
momenta; the constant $C$ contains the coupling constant,
together with numerical factors such as isospin, symmetry factors,
factors of $\pi$ from the integration measure and so on.  This will 
be made clear for specific cases discussed later.  Note that
from the three external momenta ($P, Q, K$), two of which are
independent, there are several choices for the independent Lorentz
scalar quantities that a form factor depends on.  The choice of form
factor variables will be dictated by the process in question.  The
form in Eq.~(\ref{eq:ffPVV}) facilitates a study of the dependence
upon the momenta of the vector objects.  To study the dependence upon
the momentum \mbox{$K = -(P+Q)$} of the pseudoscalar object, we would
take advantage of the antisymmetry of the $\epsilon$ tensor to use
\begin{eqnarray} 
 \Lambda^{PVV}_{\mu\nu}(-(Q+K);Q) &=& C \; 
        F_{PVV}\left(K^2,(K+Q)^2,Q^2\right) 
\nonumber \\ &&{} \times
        \epsilon_{\mu\nu\rho\sigma} \, Q_\rho \, K_\sigma ~.
\end{eqnarray}

For electromagnetic interactions, electromagnetic current conservation
is manifest if the approximations used for the dressing of the quark
propagators, meson Bethe--Salpeter amplitudes [BSAs], and the
quark-photon vertices are dynamically consistent with the
approximation used for the photon-hadron interaction.  Rainbow-ladder
truncation of the DSE and BSE, in combination with impulse
approximation for the photon-meson coupling satisfies this consistency
requirement~\cite{Maris:2000sk}.

\subsection{\label{sec:DSEs}
Dyson--Schwinger Equations}
The DSE for the renormalized quark propagator in Euclidean space is
\begin{eqnarray}
S(p)^{-1} &=& i \, Z_2\, /\!\!\!p + Z_4\,m_q(\mu) +
\nonumber \\ && {} 
        Z_1 \int_q^\Lambda\! g^2 D_{\mu\nu}(k) 
        \frac{\lambda^a}{2} \gamma_\mu S(q)\Gamma^a_\nu(q,p) \,,
\label{gendse}
\end{eqnarray}
where $D_{\mu\nu}(k)$ is the dressed-gluon propagator,
$\Gamma^a_\nu(q,p)$ the dressed-quark-gluon vertex, and \mbox{$k=p-q$}.  
The most general solution of Eq.~(\ref{gendse}) has the form
\mbox{$S(p)^{-1} = i /\!\!\! p A(p^2) +$} \mbox{$B(p^2)$} and is renormalized
at spacelike $\mu^2$ according to \mbox{$A(\mu^2)=1$} and
\mbox{$B(\mu^2)=m_q(\mu)$} with $m_q(\mu)$ being the current quark mass.  We
use the Euclidean metric where \mbox{$\{\gamma_\mu,\gamma_\nu\} =
2\delta_{\mu\nu}$}, \mbox{$\gamma_\mu^\dagger = \gamma_\mu$} and
\mbox{$a\cdot b = \sum_{i=1}^4 a_i b_i$}.

Mesons can be studied by solving the homogeneous BSE for 
$q^a \bar{q}^b$ bound states
\begin{eqnarray}
 \Gamma^{a\bar{b}}_H(p_+,p_-) &=& 
        \int_q^\Lambda\! K(p,q;P) 
\nonumber \\ && {} 
        \otimes S^a(q_+) \, \Gamma^{a\bar{b}}_H(q_+,q_-) \, S^b(q_-)\, ,
\label{homBSE}
\end{eqnarray}
where $a$ and $b$ are flavor indices, $p_+ = p + \eta P$ and $p_- = p -
(1-\eta) P$ are the outgoing and incoming quark momenta respectively,
and  $q_\pm$ is defined similarly.  The kernel $K$ is the renormalized,
amputated $q\bar q$ scattering kernel that is irreducible with respect
to a pair of $q\bar q$ lines.  This equation has solutions at discrete
values of $P^2 = -m_H^2$, where $m_H$ is the meson mass.  Together
with the canonical normalization condition for $q\bar q$ bound states,
it completely determines $\Gamma_H$, the bound state BSA.  The
different types of mesons, such as pseudo-scalar, vector, etc are
characterized by different Dirac structures.  The most general
decomposition for pseudoscalar bound states is~\cite{Maris:1997tm}
\begin{eqnarray}
\label{genpion}
\Gamma_{PS}(q_+,q_-) &=& \gamma_5 \big[ i E(q^2;q\cdot P;\eta) 
\nonumber \\ && {}
        + \;/\!\!\!\! P \, F(q^2;q\cdot P;\eta) 
        + \,/\!\!\!k \, G(q^2;q\cdot P;\eta) 
\nonumber \\ && {}
        + \sigma_{\mu\nu}\,P_\mu q_\nu \,H(q^2;q\cdot P;\eta) \big]\,,
\end{eqnarray}
where the invariant amplitudes $E$, $F$, $G$ and $H$ are Lorentz
scalar functions of $q^2$ and $q\cdot P$.  For charge eigenstates,
these amplitudes are appropriately odd or even in the charge parity
odd quantity $q\cdot P$.  In the case of the $0^{-+}$ pion, for
example, the amplitude $G$ is odd in $q\cdot P$, the others are even.
Note also that these amplitudes explicitly depend on the momentum
partitioning parameter $\eta$.  However, so long as Poincar\'e
invariance is respected, the resulting physical observables are
independent of this parameter~\cite{Maris:2000sk}.

The meson BSA $\Gamma^{a\bar{b}}$ is normalized according to the
canonical normalization condition
\begin{eqnarray}
2\,P_\mu &=& 
        N_c\frac{\partial}{\partial P_\mu} {\rm Tr}\Big\{ 
        \int^\Lambda_q \!\! \bar\Gamma^{b\bar{a}}(\tilde{q}',\tilde{q})\,
        S^a(q_+)\,\Gamma^{a\bar{b}}(\tilde{q},\tilde{q}')\,S^b(q_-)
\nonumber\\ &&
        + \int^\Lambda_{k,q} \, 
        \bar\chi^{b\bar{a}}(\tilde{k}',\tilde{k})\,K(k,q;P)\,
        \chi^{a\bar{b}}(\tilde{q},\tilde{q}') \Big\} \, ,
\label{gennorm}
\end{eqnarray}
at ${P^2=Q^2=-m_H^2}$, with \mbox{$\tilde{q} = q+$}
\mbox{$\eta Q$}, \mbox{$\tilde{q}'=q-$} \mbox{$(1-\eta) Q$}, and 
similarly for $\tilde{k}$ and $\tilde{k}'$.  For vector mesons, it is
understood that one must contract and average over the Lorentz indices
of the (transverse) BSAs to account for the three independent
polarizations.

We will later need the following exact expression for the pion decay
constant $f_\pi$:
\begin{equation}
   f_\pi P_\mu = 
        Z_2 \, N_c \!\int^\Lambda_q \!\!{\rm Tr}\big[ \gamma_5 \gamma_\mu  
                S(q_+) \Gamma_\pi(q_+, q_-) S(q_-)\big]\, .
\label{eq:fpi}
\end{equation}

Since a massive vector meson bound state is transverse, the BSA
requires eight covariants for its representation.  We choose the
transverse projection of the form
\begin{eqnarray}
\Gamma_\mu(q_+,q_-) &=& 
        \gamma_\mu \, V_1 + q_\mu\not\!q \, V_2
        + q_\mu\not\!P \, V_3
\nonumber \\   && {} 
        + \gamma_5\epsilon_{\mu\alpha\nu\beta}\gamma_\alpha 
                q_\nu P_\beta \, V_4
        + q_\mu \, V_5
\nonumber \\   && {} 
        + \sigma_{\mu\nu}q_\nu  \, V_6
        + \sigma_{\mu\nu}P_\nu \, V_7
\nonumber \\   && {} 
        + q_\mu \sigma_{\alpha\beta}q_\alpha P_\beta \, V_8 \;.
\label{vecBSAform}
\end{eqnarray}
This form is a variation of that used in Ref.~\cite{Maris:1999nt} that
is simpler and easier to use in many respects.  The invariant
amplitudes $V_i$ are Lorentz scalar functions of $q^2$ and $q\cdot P$
and again, for charge eigenstates, they are either odd or even in
$q\cdot P$.  For the $1^{--}$ $\rho$ meson, $V_3$ and $V_6$ are odd,
the other amplitudes are even.

The quark-photon vertex is \mbox{$\tilde{\Gamma}^a_\mu = \hat{Q}^a \,
\Gamma^a_\mu $} where $\hat{Q}^a$ is the $a$-quark electric charge and
the amplitude $\Gamma^a_\mu$ is normalized so that its bare (UV) limit
is $\gamma_\mu$.  The vector vertex $\Gamma^a_\mu$ with total momentum
\mbox{$Q= p_+ - p_-$} satisfies the inhomogeneous BSE
\begin{eqnarray}
 \Gamma^a_\mu(p_+,p_-) &=& Z_2 \, \gamma_\mu + 
        \int^\Lambda_q \! K(p,q;Q) 
\nonumber \\ && {}
        \otimes S^a(q_+) \, \Gamma^a_\mu(q_+,q_-)\, S^a(q_-)\, ,
\label{verBSE}
\end{eqnarray}
where we ignore the possibility of flavor mixing in the kernel.  There
are twelve invariant amplitudes needed to represent the quark-photon
vertex.  The four longitudinal ones can be expressed directly in terms
of the quark propagator amplitudes via the vector Ward--Takahashi
identity [WTI]; the eight transverse amplitudes are defined by a
decomposition of the form of Eq.~(\ref{vecBSAform}) and are obtained
from solutions of Eq.~(\ref{verBSE}) as discussed in
Ref.~\cite{Maris:1999bh}.  For timelike $Q^2$ near the position of a
vector meson bound state with mass $m_V^2$, the transverse part of the
quark-photon vertex has the resonance pole
behavior~\cite{Maris:1999bh}
\begin{equation}
 \Gamma^a_\mu(p_+,p_-) \rightarrow \frac{f_V \, m_V}{Q^2 + m_V^2} \, 
                \Gamma_\mu^{a\bar{a}\,V}(p_+,p_-) \; ,
\label{comres}
\end{equation}
where $f_V$ is the electroweak decay constant of the vector meson.

\subsection{\label{sec:model}
Model Truncation}
We employ the model that has been developed recently for an efficient
description of the masses and decay constants of the light
pseudoscalar and vector mesons~\cite{Maris:1997tm,Maris:1999nt}.  This
consists of the rainbow truncation of the DSE for the quark propagator
and the ladder truncation of the BSE for the mesons.  The required
effective $\bar q q$ interaction has a phenomenological infrared
behavior and reduces to the perturbative QCD running coupling in the
ultraviolet~\cite{Maris:1997tm,Maris:1999nt}.  In particular, the
rainbow truncation of the quark DSE, Eq.~(\ref{gendse}), is
\begin{equation}
\label{ourDSEansatz}
Z_1 \, g^2 D_{\mu \nu}(k) \Gamma^i_\nu(q,p) \rightarrow
 {\cal G}(k^2) D_{\mu\nu}^{\rm free}(k)\, \gamma_\nu
                                        \textstyle\frac{\lambda^i}{2} \,,
\end{equation}
where $D_{\mu\nu}^{\rm free}(k=p-q)$ is the free gluon propagator in
Landau gauge.  The consistent ladder truncation of the BSE,
Eq.~(\ref{homBSE}), is
\begin{equation}
\label{ourBSEansatz}
        K(p,q;P) \to
        -{\cal G}(k^2)\, D_{\mu\nu}^{\rm free}(k)
        \textstyle{\frac{\lambda^i}{2}}\gamma_\mu \otimes
        \textstyle{\frac{\lambda^i}{2}}\gamma_\nu \,,
\end{equation}
where \mbox{$k=p-q$}.  These two truncations are consistent in the
sense that the combination produces vector and axial-vector vertices
satisfying the respective WTIs.  In the axial case, this ensures that
in the chiral limit the ground state pseudoscalar mesons are the
massless Goldstone bosons associated with chiral symmetry
breaking~\cite{Maris:1997tm,Maris:1998hd}.  In the vector case, this
ensures electromagnetic current conservation if the impulse
approximation is used to describe the meson electromagnetic current or
charge form factor~\cite{Maris:2000sk}.  Furthermore, this truncation
was found to be particularly suitable for the flavor octet
pseudoscalar and vector mesons since the next-order contributions in a
quark-gluon skeleton graph expansion, have a significant amount of
cancellation between repulsive and attractive
corrections~\cite{Bender:1996bb}.

The model is completely specified once a form is chosen for the
``effective coupling'' ${\cal G}(k^2)$.  We employ the
Ansatz~\cite{Maris:1999nt}
\begin{eqnarray}
\label{gvk2}
\frac{{\cal G}(k^2)}{k^2} &=&
        \frac{4\pi^2\, D \,k^2}{\omega^6} \, {\rm e}^{-k^2/\omega^2}
\nonumber \\ && {}
        + \frac{ 4\pi^2\, \gamma_m \; {\cal F}(k^2)}
        {\textstyle{\frac{1}{2}} \ln\left[\tau + 
        \left(1 + k^2/\Lambda_{\rm QCD}^2\right)^2\right]} \;,
\end{eqnarray}
with \mbox{$\gamma_m=12/(33-2N_f)$} and \mbox{${\cal F}(s)=(1 -
\exp\frac{-s}{4 m_t^2})/s$}.  The ultraviolet behavior is chosen to be
that of the QCD running coupling $\alpha(k^2)$; the ladder-rainbow
truncation then generates the correct perturbative QCD structure of
the DSE-BSE system of equations.  The first term implements the strong
infrared enhancement in the region \mbox{$0 < k^2 < 1\,{\rm GeV}^2$}
phenomenologically required~\cite{Hawes:1998cw} to produce a realistic
value for the chiral condensate.  We use \mbox{$m_t=0.5\,{\rm GeV}$},
\mbox{$\tau={\rm e}^2-1$}, \mbox{$N_f=4$}, \mbox{$\Lambda_{\rm QCD} =
0.234\,{\rm GeV}$}, and a renormalization scale \mbox{$\mu=19\,{\rm
GeV}$} which is well into the perturbative
domain~\cite{Maris:1997tm,Maris:1999nt}.  The remaining parameters,
\mbox{$\omega = 0.4\,{\rm GeV}$} and \mbox{$D=0.93\,{\rm GeV}^2$}
along with the quark masses, are fitted to give a good description of
the chiral condensate, $m_{\pi/K}$ and $f_{\pi}$.

Within this model, the quark propagator reduces to the one-loop
perturbative QCD propagator in the ultraviolet region.  In the
infrared region both the wave function renormalization $Z(p^2) =
1/A(p^2)$ and the dynamical mass function $M(p^2)=B(p^2)/A(p^2)$
deviate significantly from the perturbative behavior, due to chiral
symmetry breaking.  Recent
comparisons~\cite{Maris:2000zf,Tandy:2001qk} of results from this
rainbow DSE model to lattice QCD
simulations~\cite{Skullerud:2000un,Skullerud:2001aw} provide
semiquantitative confirmation of the behavior generated by the present
DSE model: a significant enhancement of $M(p^2)$ and a modest
enhancement of $A(p^2)$ below $1\;{\rm GeV}^2$.

The vector meson masses and electroweak decay constants produced by
this model are in good agreement with experiments~\cite{Maris:1999nt},
as can be seen from Table~\ref{tab:masses}.
\begin{table}
\caption{\label{tab:masses}
Overview of the results of the model for the meson masses and decay
constant, adapted from Refs.~\protect\cite{Maris:1997tm,Maris:1999nt}.}
\begin{ruledtabular}
\begin{tabular}{l|ccr}
        & \multicolumn{1}{r}{experiment~\protect\cite{Groom:2000in}}
        & \multicolumn{1}{r}{calculated}  &\\
        & \multicolumn{1}{r}{(estimates)}
        & \multicolumn{1}{r}{($^\dagger$ fitted)} &\\ \hline
$m^{u=d}_{\mu=1 {\rm GeV}}$ &
        \multicolumn{1}{r}{ 5 - 10 MeV}  &
        \multicolumn{1}{r}{ 5.5 MeV}     &\\
$m^{s}_{\mu=1 {\rm GeV}}$ &
        \multicolumn{1}{r}{ 100 - 300 MeV} &
        \multicolumn{1}{r}{ 125 MeV   }    &\\ \hline
- $\langle \bar q q \rangle^0_{\mu}$
                & (0.236 GeV)$^3$ & (0.241$^\dagger$)$^3$ &\\
$m_\pi$         &  0.1385 GeV &   0.138$^\dagger$ &\\
$f_\pi$         &  0.131 GeV &   0.131$^\dagger$ &\\
$m_K$           &  0.496 GeV  &   0.497$^\dagger$ &\\
$f_K$           &  0.160 GeV  &   0.155        &\\ \hline
$m_\rho$        &  0.770 GeV  &   0.742        &\\
$f_\rho$        &  0.216 GeV  &   0.207        &\\
$m_{K^\star}$   &  0.892 GeV  &   0.936        &\\
$f_{K^\star}$   &  0.225 GeV  &   0.241        &\\
$m_\phi$        &  1.020 GeV  &   1.072        &\\
$f_\phi$        &  0.236 GeV  &   0.259        &\\
\end{tabular}
\end{ruledtabular}
\end{table}
Without any readjustment of the parameters, this model agrees
remarkably well with the most recent Jlab data~\cite{Volmer:2000ek}
for the pion charge form factor $F_\pi(Q^2)$.  Also the kaon charge
radii and electromagnetic form factors are well
described~\cite{Maris:2000sk,Maris:2000wz}.  The strong decays of the
vector mesons into a pair of pseudoscalar mesons are also
well-described within this model~\cite{Maris:2001rq,JMT01prep}.


\section{\label{sec:pigg}
The $\pi\,\gamma\,\gamma$ transition}
The symmetrized invariant amplitude for the coupling of a pion with
momentum $K=-(Q_1+Q_2)$ to a pair of photons with helicities
$\lambda_1$ and $\lambda_2$ and momenta $Q_1$ and $Q_2$ has the form
\mbox{${\cal M}^{\lambda_1\lambda_2} = \epsilon^{\lambda_1}_\mu \,
\Lambda^{P\gamma\gamma}_{\mu\nu} \,\epsilon^{\lambda_2}_\nu $} where
$\epsilon^\lambda_\mu$ is a photon polarization vector.  With
consideration of the two flavor-labeled components of neutral pion and
the two orderings of photons, the vertex can be decomposed as
\begin{eqnarray}
\label{pigg_vertex}
\Lambda^{\pi^0 \gamma\gamma}_{\mu\nu} &=& \frac{2}{\sqrt{2}} 
          \left[(\hat{Q}^u)^2 \Lambda_{\mu\nu}^u
                     - (\hat{Q}^d)^2 \Lambda_{\mu\nu}^d \right] 
\end{eqnarray}
where $\hat{Q}^a$ is the electric charge of the $a$-flavored quark.
Here $\Lambda_{\mu\nu}^a$ is the vertex contribution from a given
ordering of the photons and from flavor component $a \bar{a}$ of the
$\pi^0$.  The factor of $\sqrt{2}$ in the denominator comes from the
flavor weights in the $\pi^0$ state, $(u\bar{u} - d\bar{d})/\sqrt{2}$,
and the factor of $2$ in the numerator comes from summation over the
two orderings of the photon vertices.  Isospin symmetry gives
\mbox{$\Lambda_{\mu\nu}^u = \Lambda_{\mu\nu}^d$}; thus we have
\mbox{$\Lambda^{\pi^0 \gamma \gamma}_{\mu\nu} =
\sqrt{2}\,\Lambda_{\mu\nu}^u /3$}.  The impulse approximation result
for the vertex $\Lambda_{\mu\nu}^u$ is
\begin{eqnarray}
\Lambda^u_{\mu\nu}(Q_1;Q_2) &=& 
     e^2\, N_c \int^\Lambda_k \! {\rm Tr}\Big[
                S(q_2) \, \Gamma^\pi(q_2,q_1) \, S(q_1) 
\nonumber \\ &&  {}\times
        i\Gamma_\mu(q_1,k)\, S(k) \, i\Gamma_\nu(k,q_2) \Big] \;.
\label{eq:pigg}
\end{eqnarray}
with $q_1 = k + Q_1$ and $q_2 = k - Q_2$.  The result is independent
of the choice of integration variable, and this provides a useful
check on numerical methods.

The form factor that we associate with this process is defined
from~\cite{Frank:1995gc}
\begin{eqnarray}
\lefteqn{\Lambda^{\pi^0\gamma\gamma}_{\mu\nu}(Q_1;Q_2) = 
        \frac{\sqrt{2}}{3} \,\Lambda_{\mu\nu}^u(Q_1,Q_2) = }
\nonumber \\ 
&& = \frac{2\,i\,\alpha_{\rm em}\,g_{\pi\gamma\gamma} }{\pi\,\tilde{f_\pi}}
  \,\epsilon_{\mu \nu \rho \sigma }\,Q_{1\rho } Q_{2\sigma}
        F_{\pi\gamma\gamma}(Q_1^2,Q_2^2)  \,,
\label{eq:piggff}
\end{eqnarray}
where $\tilde{f_\pi} = f_\pi/\sqrt{2} = 92\;{\rm MeV}$ and the fine
structure constant is \mbox{$\alpha_{\rm em}= e^2/4\pi$}.  The Lorentz
scalar \mbox{$Q_1 \cdot Q_2$} does not occur as a third argument of
the form factor since the pion mass-shell constraint fixes it in terms
of those shown.  The form factor so defined has the normalization
$F_{\pi\gamma\gamma}(0,0) = 1$ so that $g_{\pi\gamma\gamma}$ is the
coupling constant.  For on-shell photons, $Q^2_1 = Q^2_2 = 0$, this
vertex describes the neutral pion decay and the associated width is
\begin{eqnarray}
 \Gamma_{\pi^0 \gamma \gamma} &=& 
        \frac{g_{\pi\gamma\gamma}^2 \alpha_{\rm em}^2 m_\pi^3}
        {16 \pi^3 \tilde{f_\pi}^2}~.
\label{piwidth}
\end{eqnarray}
This decay is governed by the axial anomaly, which leads to
$g^0_{\pi\gamma\gamma}=1/2$ in chiral limit.  For example, the
corresponding decay width produced by Eq.~(\ref{piwidth}) is
$\Gamma=7.7\;{\rm eV}$, while the experimental value is
$\Gamma=7.8\;{\rm eV}$.

A consistent ladder-rainbow truncation of the DSE-BSE system of
equations preserves the symmetry constraints that dictate this
anomaly~\cite{Roberts:1996hh,Maris:1998hc}.  Our approach indeed
reproduces the correct neutral pion decay width.  The difference in
the coupling constant produced by the finite size of $m_\pi$ is less
than 2\%, which is our estimate of our numerical accuracy.  Since the
decay width is very sensitive to $m_\pi$, we use the experimental
$m_\pi$ to convert the experimental width to the coupling constant
\mbox{$g_{\pi \gamma \gamma}^{\rm expt}=$} \mbox{$0.501 \pm 0.018$};
our theoretical result is \mbox{$g_{\pi\gamma\gamma}^{\rm th} =
0.502$}.

\subsection{\label{sec:ffpigg}
$\gamma^*\,\pi\,\gamma$ transition form factor}
\begin{figure}[ht]
\includegraphics[width=8cm]{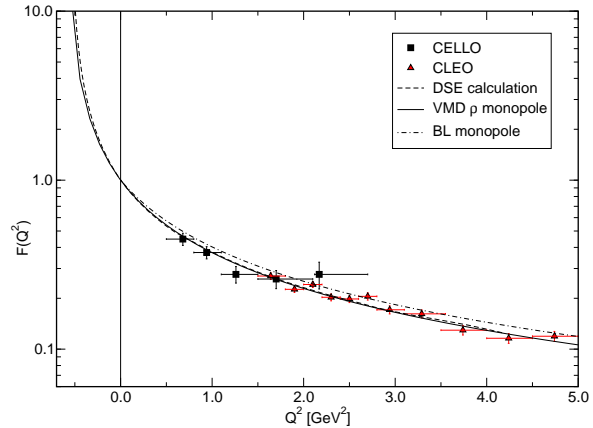}
\caption{ 
Our results for the $\gamma^* \pi \to \gamma$ form factor, together with 
a simple VMD monopole with mass scale $m_\rho^2 = 0.59\;{\rm GeV}^2$, 
a monopole based on the Brodsky--Lepage asymptotic form $0.674/(Q^2 + 
0.674)$, and data from CLEO~\protect\cite{Gronberg:1998fj} and
CELLO~\protect\cite{Behrend:1991sr}.  }
\label{fig:pigg}
\end{figure}

For one on-shell photon, we can define a transition form factor
$F_{\gamma^*\pi\gamma}(Q^2) = F_{\pi\gamma\gamma}(Q^2,0)$ for the
process $\gamma^* \pi \to \gamma$, which has been measured by the CLEO
and CELLO collaborations~\cite{Behrend:1991sr,Gronberg:1998fj}.  Our
numerical results for this form factor are presented in
Fig.~\ref{fig:pigg}, and agree very well with the available data.  The
corresponding interaction radius, defined by $r^2 = -6 F'(0)$ is
$r_{\pi\gamma\gamma}^2 = 0.39~{\rm fm}^2$.   This is in much better
agreement with the experimental estimate~\cite{Behrend:1991sr} of 
$r_{\pi\gamma\gamma}^2 = 0.42 \pm 0.04~{\rm fm}^2$ than an earlier 
theoretical result~\cite{Frank:1995gc} within the present DSE framework 
that had a considerably greater reliance upon phenomenological elements. 

Our numerical results indicate that this form factor is an almost
perfect monopole, and can be simulated very well by a VMD formula
$F(Q^2) = m_\rho^2/(Q^2 + m_\rho^2)$ with a $\rho$-meson mass scale
$m_\rho^2 = 0.59~{\rm GeV}^2$, over the entire momentum range shown.
With use of perturbative QCD and factorization on the lightfront,
Brodsky and Lepage~\cite{Lepage:1980fj} showed that this form factor
behaves like $Q^2 F(Q^2) \rightarrow 8 \pi^2 \tilde{f_\pi}^2 =
0.674~{\rm GeV}^2$ in the asymptotic spacelike region (\mbox{$Q^2 >0
$} here).  Our results indicate a slightly lower asymptotic mass
scale, see Fig.~\ref{fig:pigg}.  We will discuss the asymptotic
behavior further in Sec.~\ref{sec:piggasym} below.

\subsection{\label{sec:ffpiggsym}
Equal photon momenta}
Next, we consider the $\pi^0 \gamma^* \gamma^*$ form factor,
corresponding to an on-shell pion, and two off-shell photons that have
equal spacelike virtualities $Q^2$.  Although experimental data are
not available for this equally virtual configuration, it is
interesting from a theoretical point of view.

\begin{figure}[ht]
\includegraphics[width=8cm]{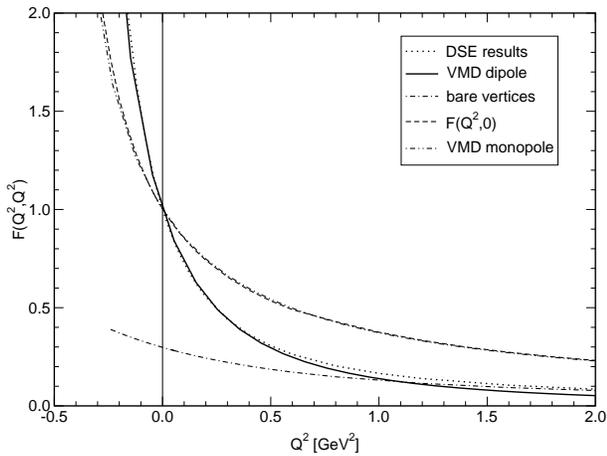}
\caption{ 
Our DSE results for the symmetric $\gamma^* \pi \to \gamma^*$ form
factor at low momentum together with a simple VMD dipole with mass
scale $m_\rho^2 = 0.59~{\rm GeV}^2$.  For comparison, the $\gamma^*
\pi \to \gamma$ form factor $F(Q^2,0)$ and the corresponding VMD
monopole from Fig.~\protect\ref{fig:pigg} are also shown.}
\label{fig:piggsym}
\end{figure} 

\begin{figure}
\includegraphics[width=8cm]{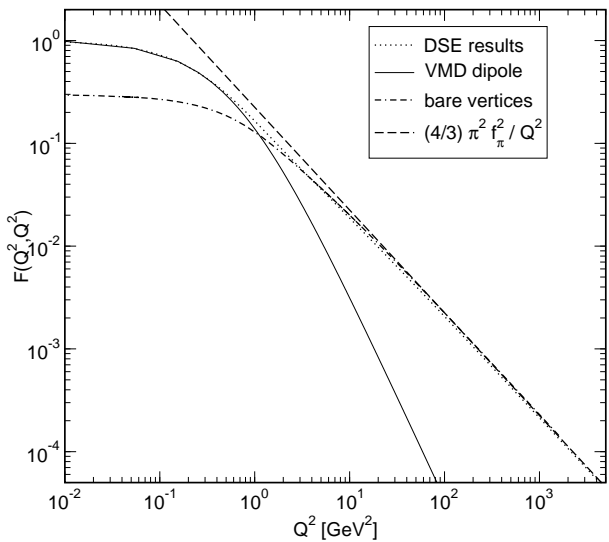}
\caption{ 
Our DSE results for the symmetric $\gamma^* \pi \to \gamma^*$ form
factor, compared to the derived asymptotic $1/Q^2$ behavior.  Note
that \mbox{$f_\pi=131~{\rm MeV}$}.  The naive VMD model suggests a
dipole behavior which is correct only in the infrared.
\label{fig:piggsymlog}}
\end{figure}

At low $Q^2$ this $\gamma^* \pi \to \gamma^*$ form factor is
reasonably well represented by a dipole, in contrast to the $\gamma^*
\pi \to \gamma$ case, which is an almost perfect monopole, see
Fig.~\ref{fig:piggsym}.  This dipole behavior is consistent with VMD;
in the symmetric case VMD leads to dipole behavior since both photons
behave like $1/(Q^2 + m_\rho^2)$ near the $\rho$-pole.  However, for
spacelike momenta of about $Q^2 > 0.5~{\rm GeV}^2$, the influence of
the vector meson resonance has weakened considerably and the behavior
becomes increasingly a monopole form.  This can be seen more clearly
in Fig.~\ref{fig:piggsymlog}.

\subsection{\label{sec:piggasym}
Asymptotic behavior}
As pointed out recently~\cite{Anikin:1999cx}, it is convenient to
discuss the asymptotic behavior of this form factor with reference to
the following form produced by the light-cone operator product
expansion~\cite{Lepage:1980fj,Chase:1980ck}
\begin{eqnarray}
 F(Q_1^2,Q_2^2) &\rightarrow&
        4 \pi^2 \tilde{f_\pi}^2\Big\{\frac{J(\omega)}{Q_1^2+Q_2^2} 
\nonumber \\ {} &&+ 
        {\cal O}\left(\frac{\alpha_s}{\pi}, \frac{1}{(Q_1^2+Q_2^2)^2}
        \right) \Big\}~.
\label{eq:gFOPE}
\end{eqnarray}
Here $\omega$ is the photon asymmetry  $(Q_1^2 - Q_2^2)/(Q_1^2 +
Q_2^2)$.  In lightfront QCD $J(\omega)$ is related to the leading twist 
pion distribution amplitude $\phi_\pi(x)$ by
\begin{eqnarray}
  J(\omega) &=& \frac{4}{3}\int_0^1 \!\!
        \frac{dx}{1-\omega^2(2x - 1)} \phi_\pi(x)~.
\end{eqnarray}
The normalization of $\phi_\pi(x)$ immediately gives $J(0) =
\frac{4}{3}$ for the case of equal photon virtuality.  Since a
corresponding result does not exist for \mbox{$\omega = 1$}, one
expects more model dependence for that asymmetric case.

Within the present DSE-based approach, it is straightforward to
analyze the asymptotic behavior for equal virtuality.  From
Eq.~(\ref{eq:pigg}) with photon momenta $Q_1 = P - K/2$ and $Q_2 = - P
- K/2$, and pion momentum $K$, we have
\begin{eqnarray}
\Lambda^u_{\mu\nu}(Q_1;Q_2) &=& 
        - e^2 N_c \int^\Lambda_q \! {\rm Tr}\Big[
                S(q_+) \, \Gamma^\pi(q_+,q_-) \, S(q_-) 
\nonumber \\ &&  {}\times
        \Gamma_\mu(q_-,\tilde{q})\, S(\tilde{q}) \, 
        \Gamma_\nu(\tilde{q},q_+) \Big] \;.
\label{eq:piggas}
\end{eqnarray}
where $q_\pm = q \pm K/2$, and $\tilde{q} = q+P$.  The pion mass-shell
condition coupled with $m_\pi$ being negligible compared to the other
scales leads to the asymptotic domain being characterized by $Q_1^2 =
Q_2^2 \simeq P^2\gg K^2 $.  In this perturbative domain, we have
\begin{eqnarray}
\lefteqn{\Gamma_\mu(q-K/2,q+P)\; S(q+P)\;\Gamma_\nu(q+P,q+K/2) \to }
\nonumber \\
&& Z_2^2 \; \gamma_\mu \; S^0(q+P)\;  \gamma_\nu
= -i\, Z_2 \, \epsilon_{\mu\nu\alpha\beta}\, \gamma_5 \gamma_\alpha 
                                                     P_\beta / P^2 
\nonumber \\
&& \;\;\;\;\;\;\;\;\;\;\;\;\;\;\;\;\;\;\;\;\;\;\;\;\;\;\;\;\;\;\;\;\;
                                \;\;\;\;\;\;\;\;\;\;\;\;\; + \cdots~,
\label{eq:gamSgam}
\end{eqnarray}
where $Z_2$ is the renormalization constant appearing in
Eq.~(\ref{gendse}) for the dressed quark propagator, and the terms not
shown make only subleading contributions to the final result.  Thus
the asymptotic behavior of the impulse approximation vertex in
Eq.~(\ref{eq:piggas}) is
\begin{eqnarray}
\Lambda^u_{\mu\nu}(Q_1;Q_2) &=&         
       i\, e^2\,Z_2\, N_c \, \epsilon_{\mu\nu\alpha\beta} \frac{P_\beta}{P^2} 
        \int^\Lambda_q \! {\rm Tr}\Big[  S(q_+)
\nonumber \\ &&
  \Gamma^\pi(q_+,q_-) \, S(q_-) 
                \gamma_5 \gamma_\alpha \Big] ~.
\label{eq:asymLambda}
\end{eqnarray}
Noting that the integral produces the pion decay constant $f_\pi$ via
Eq.~(\ref{eq:fpi}), we find from Eq.~(\ref{eq:piggff}) the asymptotic
result
\begin{eqnarray}
 F(Q_1^2,Q_2^2) \rightarrow
        \frac{16 \;\pi^2 \,\tilde{f_\pi}^2}{3\;(Q_1^2+Q_2^2)}~,
\label{eq:gF}
\end{eqnarray}
or in other words, \mbox{$J(0) = 4/3$}.   

Our numerical calculation reproduces \mbox{$J(0) = 4/3$} within
numerical accuracy as can be seen from the dashed curve in
Fig.~\ref{fig:piggsymlog} where \mbox{$Q_1^2=Q_2^2=Q^2$}.  In that
Figure we also show that, in the asymptotic region, results employing
the dressed quark-photon vertices obtained from solution of the ladder
BSE, approach rapidly the results obtained with bare vertices.  This
confirms the key approximation used to obtain
Eq.~(\ref{eq:asymLambda}) for the analytic extraction of the
asymptotic behavior within the DSE approach.

Now we return to the experimentally accessible asymmetric case of one
photon on-shell, and the other one virtual: $\omega = 1$.
Perturbative QCD on the lightfront, in combination with factorization,
suggest $J(1) = 2$~\cite{Lepage:1980fj}.  However, a monopole fit to
the experimental data, for $Q^2$ up to $10~{\rm GeV}^2$, produces
$J(1)= 1.6$~\cite{Anikin:1999cx}.  This value agrees quite well with
estimates based on a chiral quark model with a phenomenological
interaction~\cite{Anikin:1999cx} and with a fit to numerical results
of the DSE model of the present work.  Table~\ref{table:piggas}
provides a summary of these results.
\begin{table}
\caption{\label{table:piggas}
The asymptotic behavior summarized in terms of the coefficient 
$J(1)$ for the $\gamma^* \pi \gamma$ form factor, and $J(0)$
for the symmetric $\gamma^* \pi \gamma^*$ form factor, as
defined in Eq.~(\protect\ref{eq:gFOPE}).   }
\begin{ruledtabular}
\begin{tabular}{l|l}    
J(1) lightcone pQCD~\protect\cite{Lepage:1980fj,Chase:1980ck}
                                        &       2       \\
J(1) Anikin {\it et al.}~\protect\cite{Anikin:1999cx}   
                                        &       1.8     \\
J(1) earlier DSE 
analysis~\protect\cite{Klabucar:1998hr, Roberts:1998gs, Tandy:1998ha}
                                        &       4/3     \\
J(1) current DSE model numerical estimate&      1.7     \\
J(1) experimental estimate              &       1.6     \\ \hline
J(0) Anikin {\it et al.}\protect\cite{Anikin:1999cx}   
                                        &       4/3     \\
J(0) DSE analysis                       &       4/3     \\
J(0) current DSE model numerical result &     1.3     
\end{tabular}
\end{ruledtabular}
\end{table}

The asymptotic behavior of the form factor for the asymmetric case
(\mbox{$\omega =1$}) was analyzed in earlier DSE-based
works~\cite{Klabucar:1998hr, Roberts:1998gs, Tandy:1998ha}, through an
argument that is the counterpart of what we have presented in
Eqs.~(\ref{eq:asymLambda}) and (\ref{eq:gF}) for the symmetric case.
A complicating aspect of the asymmetric case in Euclidean metric is
that, because of the mass-shell constraint imposed by one photon being
real, the two photon-quark vertices and the quark propagator linking
them have momentum arguments containing imaginary parts that grow with
the asymptotic scale, but are subdominant to the real parts.  The
earlier analyses made the reasonable assumption that the leading
asymptotic behavior is produced by considering only the real arguments
(and thus the known UV limit) of the propagator and vertices; the
result obtained is \mbox{$J(1)= 4/3$}.  In the symmetric case, the
scale of the imaginary parts is dictated only by $m_\pi$, and
knowledge of the UV behavior of $\Gamma_\mu \, S \, \Gamma_\nu$ is
needed only along the real spacelike momentum axis to produce the
result given by Eq.~(\ref{eq:gF}).  We consider this situation to be
more reliable; it is confirmed by the comparison in
Fig.~\ref{fig:piggsymlog} between results with and without dressing of
the photon-quark vertices.

\section{\label{sec:rhopig}
The $\rho \,\pi\,\gamma$ and $\omega \,\pi\,\gamma$transitions}
An impulse approximation description of the $\rho^0\,\pi^0\,\gamma$,
$\rho^\pm\,\pi^\pm\,\gamma$, and $\omega^0\,\pi^0\,\gamma$ transitions
can be obtained simply by replacing one of the vector vertices in the
generic $PVV$ triangle diagram, see Eq.~(\ref{eq:genPVV}), by the
vector meson bound state BSA, and the other by the dressed
quark-photon vertex.  The only difference between these three
radiative decays is in the flavor dependent factors.

For the radiative decay \mbox{$\rho^+ \to \pi^+ \gamma$}, the photon
can radiate from either the $u$-quark or the $\bar d$-quark giving
\begin{eqnarray}
\label{rpgver}
 \Lambda^{\rho^+ \pi^+ \gamma}_{\mu \nu} &=&
        \hat{Q}_u \, \Lambda^{u\bar{d},u}_{\mu \nu}
        + \hat{Q}_{\bar{d}} \, \Lambda^{u\bar{d},\bar{d}}_{\mu \nu} \,,
\end{eqnarray}
where $\Lambda^{u\bar{d},q}_{\mu \nu}$ is the vertex having the
indicated quark flavor labeling and containing no charge or flavor
weights associated with external bosons.  With both the $\rho^0$ and
$\pi^0$ given by $(u\bar u - d\bar d)/\sqrt{2}$, the radiative decay
\mbox{$\rho^0 \to \pi^0 \gamma$} can be expressed as
\begin{eqnarray}
\label{r0p0gver}
 \Lambda^{\rho^0 \pi^0 \gamma}_{\mu \nu} &=&
        \hat{Q}_u \, \Lambda^{u\bar{u},u}_{\mu \nu}
        + \hat{Q}_{d} \, \Lambda^{d\bar{d},d}_{\mu \nu} \,,
\end{eqnarray}
where the radiative contribution from both quark and antiquark of the
same flavor have been combined.  In the isospin symmetric limit, we
have $\Lambda^{u\bar{u},u}_{\mu \nu} = \Lambda^{d\bar{d},d}_{\mu \nu}=
\Lambda^{u\bar{d},u}_{\mu \nu} = -\Lambda^{u\bar{d},\bar{d}}_{\mu\nu}$, 
and thus $\rho^\pm$ and $\rho^0$ have identical $\pi \gamma$ radiative 
decays at this level.

The radiative decay \mbox{$\omega \to \pi^0 \gamma$} is given by
\begin{eqnarray}
\label{opgver}
\Lambda^{\omega \pi \gamma}_{\mu \nu} &=&
                \hat{Q}_u \, \Lambda^{u\bar{u},u}_{\mu \nu}
                - \hat{Q}_d \, \Lambda^{d\bar{d},d}_{\mu \nu} \,,
\end{eqnarray}
where, compared to Eq.~(\ref{r0p0gver}), the change in phase of the
second term here comes from the phase of the $d \bar d$ component of
the $\omega$. Again, with isospin symmetry, we have
\mbox{$\Lambda^{\omega \pi \gamma}_{\mu \nu} =$}
\mbox{$\Lambda^{u\bar{u},u}_{\mu \nu}$} and hence 
\mbox{$\Lambda^{\omega \pi \gamma}_{\mu \nu} =$} 
\mbox{$3 \,\Lambda^{\rho \pi \gamma}_{\mu \nu}$}.

In impulse approximation, the physical vertex is given by the quark
loop integral
\begin{eqnarray}
\Lambda^{\rho \pi \gamma}_{\mu\nu}(P;Q) &=&
      e\, \frac{N_c}{3} \int^\Lambda_k \!{\rm Tr} \Big[ 
        S(q_2) \, \Gamma^\pi(q_2,q_1) \, S(q_1)  
\nonumber \\ && {} \times
        \Gamma^\rho_\mu(q_1,k)\, S(k) \,
        i \Gamma^\gamma_\nu(k,q_2) \Big] \;,
\label{eq:rhopig}
\end{eqnarray}
where $P$ is the $\rho$ momentum, the photon momentum is $Q$, and the
pion momentum is \mbox{$K= -(P+Q)$}.  We have used \mbox{$q_1 = k +
P$} and \mbox{$q_2 = k - Q$}.  This is completely analogous to
Eq.~(\ref{eq:pigg}) for the $\pi\gamma\gamma$ vertex if the
quark-photon vertex $i\Gamma_\mu$ is replaced by the $\rho$ BSA.  In
fact, if that photon momenta is continued into the timelike region,
use of Eq.~(\ref{comres}) to extract the $\rho^0$ pole contribution to
the $\pi\gamma\gamma$ vertex will identify Eq.~(\ref{eq:rhopig}) as
the $\rho\pi\gamma$ vertex.

The $\rho\pi\gamma^*$ transition form factor is identified from
the $\rho\pi\gamma$ vertex according to
\begin{equation}
\label{eq:rhopigff}
 \Lambda^{\rho \pi \gamma}_{\mu\nu}(P;Q)
        = \frac{e\, g_{\rho \pi\gamma} }{m_{\rho}}
        \, \epsilon_{\mu \nu \alpha \beta } \, P_{\alpha }Q_{\beta }
        \, F_{\rho\pi\gamma}(Q^2) \,,
\end{equation}
at $P^2 = -m_\rho^2$.  At the photon point we have \mbox{$F(0)=1$} and
$g_{\rho\pi\gamma}$ is the conventional coupling constant associated
with the radiative decay width
\begin{equation}
\Gamma_{\rho \to \pi \gamma} = 
        \frac{\alpha_{\rm em} \;g_{\rho \pi \gamma}^2}{24}
        \; m_\rho \left(1 - \frac{m_\pi^2}{m_\rho^2} \right)^3 \;.
\label{rhoradwidth}
\end{equation}
The corresponding formula holds for the $\omega \to \pi\gamma$ decay 
with \mbox{$g_{\omega \pi\gamma}/m_\omega = 
3\,g_{\rho\pi\gamma}/m_\rho$}.

As Eq.~(\ref{eq:rhopigff}) shows, it is $g_{\rho\pi\gamma}/m_\rho$
that is the natural outcome of calculations of the vector radiative
decays; therefore, it is this combination that we report in
Table~\ref{table:rhopig}.  We also give the corresponding decay
widths.  Except for the difference between the decay of the neutral
and charged states of the $\rho$, which evidently is beyond the reach
of the isospin symmetric impulse approximation, the agreement between
theory and experiment is within 10\%.  This is consistent with the
other $\pi$ and $\rho$ observables obtained from the same model.  Note
that part of the difference between the experimental decay width and
our calculated decay width comes from the phase space factor because
our calculated $\rho$ and $\omega$ masses are about 5\% too low.

\begin{table}
\caption{\label{table:rhopig}
Coupling constants and partial widths for radiative decays of $\rho$
and $\omega$, compared to experimental data~\protect\cite{Groom:2000in}.}
\begin{ruledtabular}
\begin{tabular}{l|lc|lc}        
        & calc. $g/m$           & expt. $g/m$
                        & calc. $\Gamma$& expt. $\Gamma$\\ \hline
$\rho^0 \to \pi^0\gamma$        
        &$0.68~{\rm GeV^{-1}}$  & $0.9 \pm .2$          
                        &$52~{\rm keV}$ & $102 \pm 25$  \\
$\rho^\pm \to \pi^\pm\gamma$    & $0.68$        & $0.74 \pm .05$
                        & $52$          & $68 \pm 7$    \\
$\omega^0\to\pi^0\gamma$        & $2.07$        & $2.31 \pm .08$
                        & $479$                 & $717 \pm 43$  \\
\end{tabular}
\end{ruledtabular}
\end{table}
  
\subsection{\label{sec:ffrhopig}
Transition form factor}

\begin{figure}[ht]
\includegraphics[width=8cm]{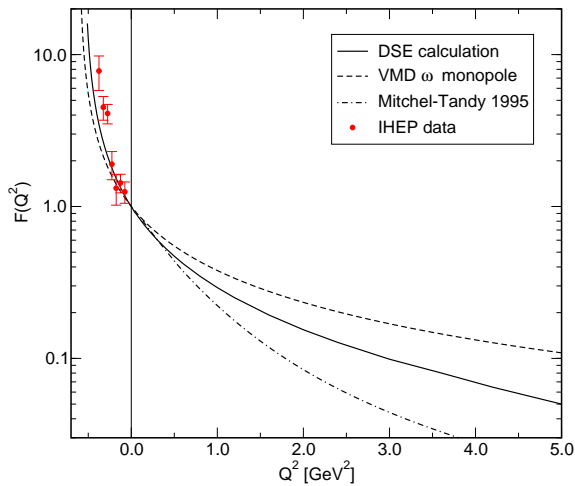}
\caption{
Our DSE result (solid line) for the $\omega\pi\gamma^*$ 
(and $\rho\pi\gamma^*$) form factor, together with a simple VMD 
monopole with mass scale $m_\omega^2 = 0.61\;{\rm GeV}^2$ (dashed line)  
and the more phenomenological result from Ref.~\cite{Tandy:1996sq}.  
The data in the timelike region are from 
Ref.~\protect\cite{Viktorov:1981jq}.  }
\label{fig:rhopig}
\end{figure}

In Fig.~\ref{fig:rhopig} we show our DSE result for the
$\omega\pi\gamma^*$ transition form factor as a function of the photon
momentum $Q^2$.  The $\rho\pi\gamma^*$ form factor is identical to
this.  Our result compares reasonably well with available experimental
data for the $\omega\pi\gamma^*$ form factor in the timelike
region~\cite{Viktorov:1981jq}; we are not aware of data in the
spacelike region.  Also shown in Fig.~\ref{fig:rhopig} is a result
from an earlier study~\cite{Tandy:1996sq} that implemented
phenomenology at the level of the dressed quark propagator and the
$\rho/\omega$ BSA.  For reference we also display the simple VMD
monopole having mass scale $m_\omega$.

Note that in the timelike region, the three different curves and the
experimental data are all very close to each other: they are dominated
by the resonance pole at the $\rho/\omega$ mass.  However, the three
curves differ significantly in the spacelike region where a monopole
clearly does not represent the behavior obtained from modeling at
quark-gluon level.  Our results can be represented, in the manner of
a Pad\'e approximant, by the formula
\begin{eqnarray}
  F_{\rho\pi\gamma}(Q^2)  &=& \frac{1.0 + Q^2}
        {1.0 + 3.04 Q^2 + 2.42 Q^4 + 0.36 Q^6} \,,
\label{eq:rhopigfit}
\end{eqnarray}
which is almost indistinguishable from our form factor on the domain
$-0.5 < Q^2 < 5~{\rm GeV}^2$.  This suggests that the asymptotic
behavior of this form factor is $1/Q^4$.  Although there are no
experimental data available for the $\rho\pi\gamma^*$ process at
spacelike photon momenta, indirect information about this form factor
can be obtained from analyses of electron scattering from hadronic
targets: e.g. it contributes as a meson exchange current tied to boson
exchange models of the nuclear interaction.  In this case the ``$\rho
\pi$ current'' is virtual and the relationship to processes involving
true quark-gluon bound states is not entirely clear.

Thus the $\rho\pi\gamma^*$ form factor plays a role in the
interpretation of electron scattering data from light nuclei, because
the isoscalar meson-exchange current contributes significantly to
these processes.  In particular our understanding of the deuteron EM
structure functions for \mbox{$Q^2 \approx 2-6~{\rm GeV}^2$} requires
knowledge of this form factor~\cite{VanOrden:1995eg}.  An initial
exploratory study~\cite{Tandy:1996sq} of the $\rho\pi\gamma^*$ vertex
within the present framework, but employing phenomenology to a much
greater extent, produced a very soft result for the form factor and
this is shown in Fig.~\ref{fig:rhopig} as the Mitchell--Tandy curve.
It was found~\cite{VanOrden:1995eg} that the resulting meson exchange
current contribution provided a very good description of the elastic
deuteron electromagnetic form factors $A(Q^2)$ and $B(Q^2)$ in the
range \mbox{$2-6~{\rm GeV}^2$} where such effects are important.

The model under consideration in the present work, based on the DSEs
of QCD, has no free parameter other than the two set by $f_\pi$ and
$\langle q \bar{q} \rangle$ as described earlier; the amount of
phenomenology is significantly less than the earlier Mitchell--Tandy
result~\cite{Tandy:1996sq}.  The present work produces a form factor
that is much softer than what is inferred from VMD but obviously
harder than the Mitchell--Tandy result~\cite{Tandy:1996sq}, as can be
seen in Fig.~\ref{fig:rhopig}.  We expect the present impulse
approximation results for the form factor to be as reliable as the
pion and kaon charge form factors from this model~\cite{Maris:2000sk}.
The present form factor is harder than what is required to fit
electron scattering from the deuteron if the standard analysis is
employed~\cite{VOrdenPrivComm}.  What is required for progress is some
estimation of the likely effect from the virtuality of the $\pi$ and
$\rho$ $\bar{q} q$ correlations relevant to the meson exchange current
mechanism.  An investigation of a related aspect of this issue is to
be found below.

\subsection{\label{sec:offshellvec}
Virtual vector $\bar{q} q$ correlations and VMD}
In Sec.~\ref{sec:ffpigg} we have seen that the $\gamma^* \to
\pi\gamma$ form factor is an almost perfect monopole with $m_\rho$ as
the mass scale, see Fig.~\ref{fig:pigg}.  This suggests that the
vector meson resonance pole in the dressed $\bar{q} \gamma^* q$ vertex
dominates to the extent that the following VMD mechanism is very
effective: there is a $\gamma$-$\rho$ transition, a $\rho$
propagation, then a $\rho$-$\pi$-$\gamma$ transition.  Implicit in
this three-step mechanism, are the assumptions: that even for
spacelike momentum of the $\gamma^*$, the first and third steps are
described by coupling constants defined at the $\rho$ mass-shell, and
the momentum dependence is carried totally by the propagator of a
point particle having the $\rho$ mass.  However, at the quark-gluon
level, one cannot define a $\rho$-meson bound state at spacelike
momentum; only a vector $\bar{q} q$ correlation with the same quantum
numbers of the $\rho$ can be discussed.  Hence the issue is the domain
of applicability and the accuracy of those assumptions.  We wish to
explore this for a variety of PVV processes; we will begin with the
connection between the $\gamma^* \,\pi\gamma$ process and the
$\rho\,\pi\gamma$ process.

We begin with Eq.~(\ref{eq:pigg}) for the quark loop integral which
produces the vertex $\Lambda^{\pi^0\gamma\gamma}_{\mu\nu}$ via
Eq.~(\ref{eq:piggff}), use the notation change \mbox{$Q_1 \to P$} and
\mbox{$Q_2 \to Q$}, and consider the domain $P^2 \approx - m_\rho^2$.
The quark-photon vertex $\Gamma_\mu(q_1,k)$ in Eq.~(\ref{eq:pigg})
then behaves like
\begin{equation}
 \Gamma_\mu(q_1,k) \sim
        \frac{f_\rho \,m_\rho}{P^2 + m_\rho^2} \, 
        \Gamma_\mu^\rho(q_1,k) \; .
\label{eq:gammaverpole}
\end{equation}
Use of only this resonance contribution gives 
\begin{eqnarray}
\lefteqn{\Lambda^{\pi^0\gamma\gamma}_{\mu\nu}(P;Q) \sim 
        \frac{f_\rho \, m_\rho}{P^2 + m_\rho^2} \; 
        \frac{\sqrt{2}\,e^2\, N_c}{3} }
\nonumber \\
\lefteqn{\times\!\int^\Lambda_k \!\! {\rm Tr}\Big[
        S(q_2) \Gamma^\pi(q_2,q_1) S(q_1) i\Gamma_\mu^\rho(q_1,k) 
                                        S(k) i\Gamma_\nu(k,q_2) \Big] }
\nonumber \\ &=& \frac{i\,e\,\sqrt{2}\;f_\rho m_\rho}{P^2 + m_\rho^2} \;
        \Lambda^{\rho \pi \gamma}_{\mu\nu}(P;Q) \;,
\;\;\;\;\;\;\;\;\;\;\;\;\;\;\;\;\;\;\;\;\;\;\;\;\;\;\;\;
\end{eqnarray}
where the last equality follows from identifying the quark loop
integral for the $\rho \pi \gamma$ vertex via Eq.~(\ref{eq:rhopig}).
We substitute the general forms of both vertices from
Eqs.~(\ref{eq:piggff}) and (\ref{eq:rhopigff}) and cancel common
factors.  This yields
\begin{equation}
 \frac{g_{\pi\gamma\gamma}}{4\pi^2 \tilde{f_\pi}} \;
        F_{\gamma^*\pi\gamma^*}(P^2,Q^2) \sim 
        \frac{f_\rho \, g_{\rho\pi\gamma} }{\sqrt{2}\, (P^2 + m_\rho^2)} \;
        F_{\rho\pi\gamma^*}(Q^2)\,.
\label{eq:vertex pole}
\end{equation}
When the non-resonant photon is real, Fig.~\ref{fig:pigg} shows that
$F_{\gamma^*\pi\gamma}(P^2, Q^2=0)$ is very accurately described by
this monopole shape even for spacelike momentum as large as \mbox{$P^2
= 5\;{\rm GeV}^2 \simeq 10 m_\rho^2$}.  At \mbox{$P^2=0$}
Eq.~(\ref{eq:vertex pole}) accurately accounts for the transition
strength $g_{\rho\pi\gamma}$ in terms of $g_{\pi\gamma\gamma}$ and the
other more fundamental quantities.  The relation
\begin{eqnarray}
 \frac{g_{\pi\gamma\gamma}}{4\pi^2 \tilde{f_\pi}} 
        &\sim& \frac{f_\rho \; g_{\rho\pi\gamma}}{\sqrt{2} \; m_\rho^2 }~,
\end{eqnarray}
is borne out to the extent that substitution of the separately
calculated values of these quantities gives $0.138$ for the LHS and
$0.134$ for the RHS.

In the $\gamma^* \to \pi\gamma$ process, we can view the $\pi\gamma$
transition current as coupling to a vector $\bar{q}q$ correlation
initiated by the $\gamma^*$.  In the process $\gamma^* N \to \pi N$
away from the $s$-channel resonance region, one often considers soft
$t$-channel mechanisms modeled by meson exchange.  In the case of a
vector $t$-channel mechanism, the nucleon current is required to
provide a spacelike vector $\bar{q}q$ correlation to the $\gamma \pi$
transition current.  If we model the vector $\bar{q}q$ vertex provided
by the nucleon current as $\bar{q}(p_+) \gamma_\mu g_{VNN}(P^2)
q(p_-)$, where $g_{VNN}(P^2)$ is the phenomenological coupling
constant and form factor corresponding to the $t$-channel momentum
$P$, then we have a point $\bar{q}q$ correlation that can interact and
propagate to the $\gamma \pi$ transition current.  The object with
momentum \mbox{$P= p_+ - p_-$} that describes this satisfies the
inhomogeneous BSE
\begin{eqnarray}
 \Gamma^{VNN}_\mu(p_+,p_-) &=& Z_2 \, \gamma_\mu \; g_{VNN}(P^2) + 
        \int^\Lambda_q \! K(p,q;P) 
\nonumber \\ && {}
        \otimes S(q_+) \, \Gamma^{VNN}_\mu(q_+,q_-)\, S(q_-)\, .
\label{vcorrBSE}
\end{eqnarray}
This is just $g_{VNN}(P^2)$ times the dressed vertex seeded by
$\gamma_\mu$; i.e., $g_{VNN}(P^2)$ times the photon vertex
$\Gamma_\mu$.  Since $\Gamma_\mu$ has the vector meson pole, see
Eq.~(\ref{eq:gammaverpole}), the meson exchange mechanism, including
the ``meson'' propagation, is embedded in this vertex.

A more exact treatment would be to calculate a distributed vector
$\bar{q}q$ vertex $\bar{q}(p_+) {\cal V}_\mu(p_+,p_-)q(p_-)$ from a
quark-gluon model of the nucleon.  In the absence of a reliable
calculation of such a distributed vertex, we here explore the
consequences of the above modeling of the nucleon current as a
pointlike current $\bar{q}(p_+) \gamma_\mu g_{VNN}(P^2) q(p_-)$.  For
simplicity, we omit the phenomenological nucleon coupling strength
$g_{VNN}(P^2)$ in the calculations below, which can be easily appended
for subsequent applications such as $\gamma^* N \to \pi N$ processes.
Thus, for the $V^*\pi\gamma^*$ vertex we are led to the impulse
approximation
\begin{eqnarray}
\Lambda^{V\pi\gamma}_{\mu\nu}(P;Q) &=&
      e\, \frac{N_c}{3} \int^\Lambda_k \!{\rm Tr} \Big[ 
        S(q_2) \, \Gamma^\pi(q_2,q_1) \, S(q_1)  
\nonumber \\ && {} \times
        \Gamma^V_\mu(q_1,k)\, S(k) \,
        i \Gamma^\gamma_\nu(k,q_2) \Big] \;,
\label{eq:Vpig}
\end{eqnarray}
where $Q$ is the photon momentum, the vector current momentum is $P$,
and the pion momentum is \mbox{$K= -(P+Q)$}.  We have used
\mbox{$q_1 = k + P$} and \mbox{$q_2 = k - Q$} in complete analogy 
to Eq.~(\ref{eq:rhopig}) for the $\rho\pi\gamma$ vertex but with
$\Gamma^V_\mu$ replacing the $\rho$ BSA.  In analogy with
Eq.~(\ref{eq:rhopigff}), the general form may be expressed as
\begin{equation}
\label{eq:Vstarpigform}
 \Lambda^{V\pi\gamma}_{\mu\nu}(P;Q)
        = \frac{e\, g_{\rho \pi\gamma} }{m_{\rho} }
        \, \epsilon_{\mu \nu \alpha \beta } \, P_{\alpha }Q_{\beta }
        \,A_{V^*\pi\gamma^*}(P^2,Q^2) \,,
\end{equation}
thereby defining an amplitude $A_{V^*\pi\gamma^*}$.  Due to
Eq.~(\ref{eq:gammaverpole}), this amplitude has the physical vector
meson resonance pole at the mass-shell \mbox{$P^2 = -m_\rho^2$}.  We
thus may write for any momentum
\begin{equation}
\label{eq:rhostarpigff}
A_{V^*\pi\gamma^*}(P^2,Q^2) =
        \frac{f_\rho \, m_\rho}{P^2 + m_\rho^2}\, 
           G_{\rho^*\pi\gamma^*}(P^2,Q^2) \,,
\end{equation}
and this serves to define a generalized form factor
$G_{\rho^*\pi\gamma^*}$ for the ``process'' \mbox{$\gamma^*\rho^* \to
\pi$}.  At the $\rho$ mass-shell, this form factor
$G_{\rho^*\pi\gamma^*}$ reduces to the previously defined form factor
$F_{\rho\pi\gamma^*}(Q^2)$.  Note that this form factor is not unique
in the sense that we have chosen one particular description of the
vector $\bar{q}q$ correlation.  A different choice for the
inhomogeneous term in Eq. (\ref{vcorrBSE}) would lead to a different
amplitude $A_{V^*\pi\gamma^*}$ and thus to a different form factor
$G_{\rho^*\pi\gamma^*}$; only for on-shell $\rho$ and $\pi$ mesons can
one uniquely define a form factor $F_{\rho\pi\gamma^*}$.

Our choice for the inhomogeneous term leads to a momentum dependence
of the amplitude $A_{V^*\pi\gamma^*}$ that is the same as that of the
$\gamma^* \pi \gamma^* $ form factor, apart from the overall scale.
Hence the earlier observations associated with Fig.~\ref{fig:pigg}
indicate that, for $Q^2 \approx 0$ the monopole shape in $P^2$ with
mass scale $m_\rho$ dominates.  Thus the form factor
$G_{\rho^*\pi\gamma^*}(P^2,Q^2\simeq 0)$ in
Eq.~(\ref{eq:rhostarpigff}) has very little $P^2$ dependence and the
meson exchange picture is thus very effective there.  In other words,
the strict VMD approximation to Eq.~(\ref{eq:rhostarpigff})
\begin{equation}
\label{eq:rhostarpimexch}
A_{V^*\pi\gamma^*}(P^2,Q^2) \approx
        \frac{f_\rho \, m_\rho}{P^2 + m_\rho^2}\, 
        F_{\rho\pi\gamma^*}(Q^2)\,,
\end{equation}
is valid for $Q^2 \simeq 0$ and for the complete range of spacelike
$P^2$ investigated here ($< 5~{\rm GeV}^2$).

The question arises: what is the more general domain of $Q^2$ and
$P^2$ for the applicability of the meson exchange picture?  Certainly
when $Q^2$ and $P^2$ are both large, we can conclude from
Fig.~\ref{fig:piggsymlog} that the amplitude
$A_{V^*\pi\gamma^*}(P^2,Q^2)$ with $Q^2=P^2$ falls off like $1/P^2$,
whereas a meson dominance model would predict $1/P^4$.  Thus, the form
factor $G_{\rho^*\pi\gamma^*} (P^2,P^2)$ becomes a constant for large
$P^2$, in contrast to $F_{\rho\pi\gamma^*}(P^2)$ which seems to fall
off as $1/P^4$, as can be seen from Fig.~\ref{fig:rhopig} and our fit,
Eq.~(\ref{eq:rhopigfit}).

\begin{figure}[ht]
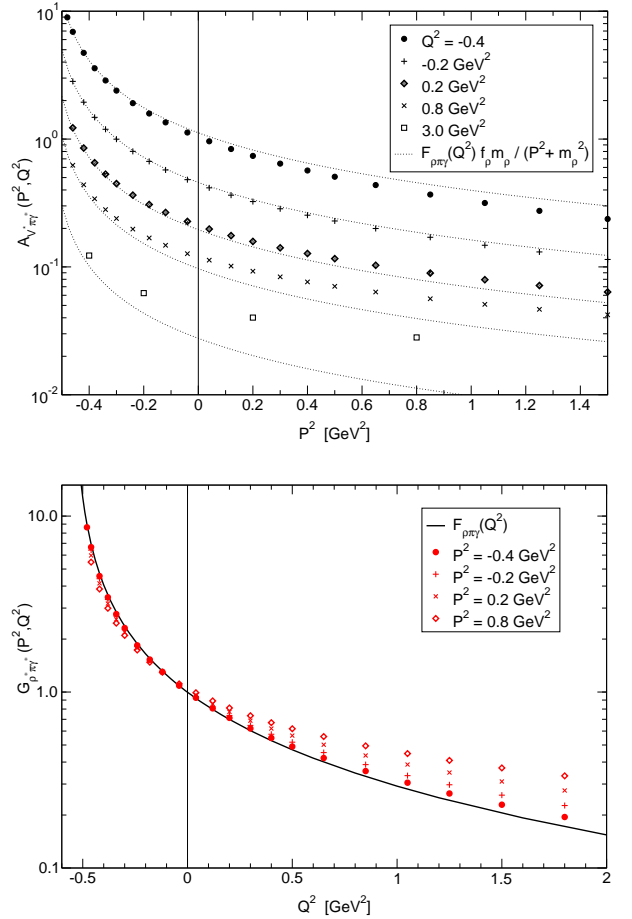

\includegraphics[width=8cm]{rhooffP2}

\vspace*{5mm}
\includegraphics[width=8cm]{rhooffQ2}
\caption{
Top: The virtual vector momentum $P^2$-dependence of the amplitude 
$A_{V^*\pi\gamma^*}$ from the DSE calculation for several values of 
the photon momentum $Q^2$ is shown by the discrete points.  The meson 
exchange approximation is shown by the curves.  
Bottom: The $Q^2$-dependence of the generalized form factor 
$G_{\rho^*\pi\gamma^*}(P^2,Q^2)$  for several values of the virtual 
vector momentum $P^2$ is shown by the discrete points.  The curve is 
the on-shell $F_{\rho\pi\gamma^*}(Q^2)$.  
\label{fig:ffinhomvec} }
\end{figure} 

In the top panel of Fig.~\ref{fig:ffinhomvec} we display as discrete
data points our results for the $P^2$-dependence of the amplitude
$A_{V^*\pi\gamma^*}(P^2,Q^2)$ for several values of the photon
momentum $Q^2$, and compare it with the naive VMD formula,
Eq.~(\ref{eq:rhostarpimexch}).  For photon momenta $Q^2$ near the
$\rho$-pole, the behavior of the amplitude $A$ as function of $P^2$ is
softer than one would expect from VMD.  This is no surprise, since we
have already seen in the previous section that the on-shell
$F_{\rho\pi\gamma^*}$ is softer than VMD.  For $Q^2 \approx 0$, the
VMD prediction works quite well.  For increasing spacelike $Q^2$ the
amplitude $A_{V^*\pi\gamma^*}$ is progressively harder in $P^2$ (i.e.
falls off slower) compared to the VMD approximation.  This means that
the $1/(P^2+ m_\rho^2)$ behavior of the $\rho$-meson propagator
increasingly over-estimates the falloff with momentum of the virtual
vector object; the meson exchange picture is showing its limitations.
As a consequence, the momentum dependence of the generalized form
factor $G_{\rho^*\pi\gamma^*}(P^2,Q^2)$ becomes more evident; compared
to the on-shell form factor $F_{\rho\pi\gamma^*}(Q^2)$ it becomes
progressively harder in $Q^2$ for increasing (spacelike) $P^2$.  This
can be seen more clearly in the bottom panel of
Fig.~\ref{fig:ffinhomvec}.

If $Q^2$ and $P^2$ are both more spacelike than about $1~{\rm GeV}^2$,
the point meson propagator in the VMD or meson exchange approximation
represented by Eq.~(\ref{eq:rhostarpimexch}) can overestimate the
falloff by 50\% or more.  We anticipate this inadequacy of the point
meson propagator to apply generally to meson exchange models.  No
standard phenomenological form factors for the coupling of virtual
"mesons" to hadronic currents (such as $g_{VNN}(P^2)$) can compensate
for the propagator falling off too fast.  In the present specific
case, we have a spacelike $\bar{q}\gamma_\mu q$ correlation, not a
"virtual bound state meson".  This $\bar{q}\gamma_\mu q$ correlation
falls off slower than a point meson propagator; in the UV region it
does not vanish, but goes to a bare vertex, $\gamma_\mu$.  Our modeling 
of the $\bar{q}\gamma_\mu q$ correlation by a ladder BSE with a point
inhomogeneous term is in agreement with perturbative QCD, whereas the
notion of a highly virtual $\rho$-meson "bound state" does not make
any sense in the perturbative region.

\subsection{\label{sec:offshellps}
Virtual pseudoscalar $\bar{q} q$ correlations}
As an extension of these observations, we also calculate the
corresponding amplitude associated with the \mbox{$\rho P^*\gamma^*$}
vertex, where $P$ stands for the pointlike pseudoscalar current
$\bar{q}\gamma_5 q$.  The interaction and propagation of this
correlation generates the pion pole.  The quantity thus made available
to us is relevant to the $\gamma^* N \to \rho N$ process via
$t$-channel exchange of a pseudoscalar $\bar{q}q$ correlation.  Of
course, in a more realistic calculation, one should use a distributed
vertex $\bar{q}(p_+) {\cal P}(p_+,p_-)q(p_-)$ from a quark-gluon model
of the nucleon, rather than a pointlike pseudoscalar vertex.
Corresponding to Eq.~(\ref{vcorrBSE}) for the vector case, our
approximation for the pseudoscalar $\bar{q}q$ correlation (with the
phenomenological $g_{PNN}(K^2)$ removed) is the inhomogeneous BSE for
the vertex
\begin{eqnarray}
 \Gamma_5(p_+,p_-) &=& Z_4 \, \gamma_5 + 
        \int^\Lambda_q \! K(p,q;Q) 
\nonumber \\ && {}
        \otimes S(q_+) \, \Gamma_5(q_+,q_-)\, S(q_-)\, ,
\label{PSverBSE}
\end{eqnarray}
where the total momentum is \mbox{$K=p_+ - p_-$}. Note that this
dressed vertex is renormalization-point {\em dependent}; however, the
combination $m_q(\mu) \Gamma_5$, where $m_q(\mu)$ is the current quark
mass, is renormalization-point independent.  Near the pion pole, this
dressed vertex $\Gamma_5$ behaves like
\begin{equation}
 \Gamma_5(p_+,p_-) \sim 
        \frac{r_P(\mu)}{K^2 + m_\pi^2} \; \Gamma^\pi(p_+,p_-)
\end{equation}
where $r_P(\mu)$ is the renormalization-point-dependent residue
in the pseudoscalar channel and is given by~\cite{Maris:1998hd}
\begin{equation}
 r_P(\mu) = Z_4 \, N_c \int^\Lambda_q \!\!
        {\rm Tr}\big[ \gamma_5 \, S(q_+) \, 
        \Gamma^\pi(q_+,q_-)\, S(q_-) \big] \, .
\end{equation}
The axial-vector WTI dictates that this pseudoscalar residue is
related to the pion mass and decay constant
through~\cite{Maris:1998hd}
\begin{eqnarray}
        2 \, m_q(\mu) \, r_P(\mu) &=& f_\pi \, m_\pi^2
\end{eqnarray}
where the renormalization-point dependence of the current quark mass
and that of the residue is such that the combination is
renormalization-point independent.

In a way that is completely parallel to Eq.~(\ref{eq:Vpig}), we
model the $\rho P^*\gamma^*$ vertex through the impulse approximation
\begin{eqnarray}
 \Lambda^{\rho P\gamma}_{\mu\nu}(P;Q) &=&
        e\, \frac{N_c}{3} \int^\Lambda_q \!{\rm Tr} \Big[ 
        S(q_2) \, \Gamma_5(q_2,q_1) \, S(q_1)  
\nonumber \\ && {} \times
        \Gamma^\rho_\mu(q_1,k)\, S(k) \,
        i \Gamma^\gamma_\nu(k,q_2) \Big] \;,
\label{eq:Prhog}
\end{eqnarray}
and note that this is renormalization-point dependent.
We can define a renormalization-point dependent amplitude 
$A_{\rho P^*\gamma^*}(K^2, Q^2)$ via
\begin{eqnarray}
\lefteqn{ \Lambda^{\rho P\gamma}_{\mu\nu}(P;Q) = }
\nonumber \\ &&
        e \, \frac{g_{\rho\pi\gamma}}{m_\rho} 
        \, \epsilon_{\mu \nu \alpha \beta } \, P_{\alpha }Q_{\beta }
        \, A_{\rho P^*\gamma^*}(K^2,Q^2) \,,
\end{eqnarray}
where $K = -(P+Q)$ is the pseudoscalar momentum and $Q$ is the photon
momentum, both incoming to the diagram.  This amplitude $A_{\rho
P^*\gamma^*}$ has the physical pion resonance pole at the mass-shell
\mbox{$K^2 = -m_\pi^2$}.  We thus may write for any momentum
\begin{eqnarray}
\label{eq:Pstarrhogff}
A_{\rho P^*\gamma^*}(K^2,Q^2) & = & 
        \frac{r_P(\mu)}{K^2 + m_\pi^2}\, 
            G_{\rho\pi^*\gamma^*}(K^2,Q^2)~,
\end{eqnarray}
and this serves to define a generalized form factor 
$G_{\rho\pi^*\gamma^*}$ for the ``process'' 
\mbox{$\gamma^* P^* \to \rho$}.  This dimensionless form factor is
renormalization-point independent, and reduces to the on-shell form
factor $F_{\rho\pi\gamma^*}(Q^2)$ at the pion mass-shell $K^2 =
-m_\pi^2$.  With $G_{\rho\pi^*\gamma^*}$ replaced by its value
at the pion mass-shell, Eq.~(\ref{eq:Pstarrhogff}) yields the meson
exchange approximation
\begin{equation}
\label{eq:Pstarrhomexch}
A_{\rho P^*\gamma^*}(K^2,Q^2) \approx
\frac{r_P(\mu)}{K^2 + m_\pi^2}\, 
F_{\rho\pi\gamma^*}(Q^2)~.
\end{equation}

\begin{figure}[ht]
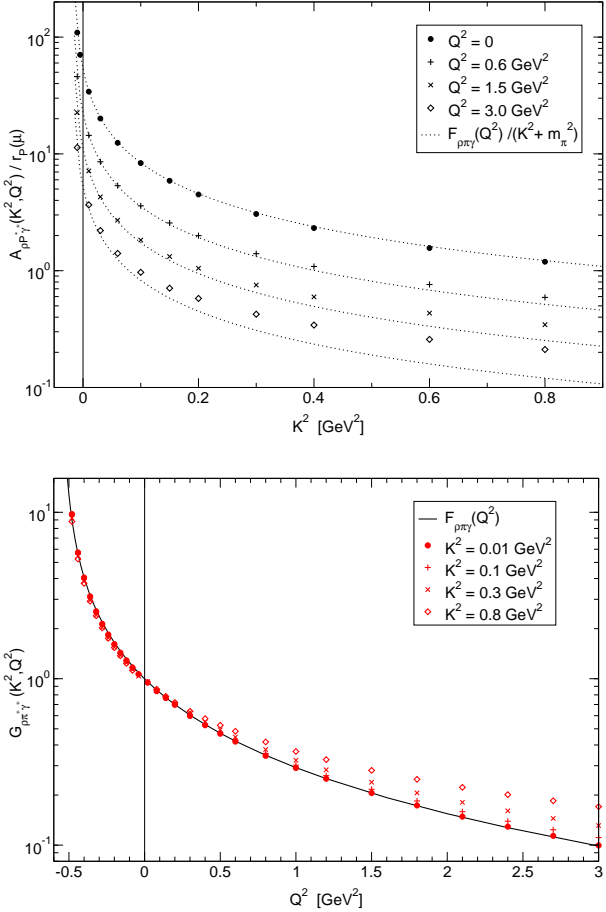

\includegraphics[width=8cm]{pionoffP2}

\vspace*{5mm}
\includegraphics[width=8cm]{pionoffQ2}
\caption{ 
Top: The dependence of the amplitude $A_{\rho P^*\gamma^*}$ on the
pseudoscalar momentum $K^2$ for several values of the photon momentum
$Q^2$ is given by the discrete points from the DSE calculation.  The
meson exchange approximation is given by the curves.
Bottom: The DSE results for the $Q^2$-dependence of the generalized
form factor $G_{\rho\pi^*\gamma^*}(K^2,Q^2)$ for several values of 
the pseudoscalar momentum $K^2$ are the discrete points.  The on-shell
$F_{\rho\pi\gamma^*}(Q^2)$ is given by the curve.
\label{fig:ffinhomps} }
\end{figure} 

The directly calculated amplitude $A_{\rho P^*\gamma^*}$ is shown in
the upper panel of Fig.~\ref{fig:ffinhomps} by the discrete points as
a function of $K^2$ for several values of $Q^2$.  The continuous
curves illustrate the meson exchange approximation,
Eq.~(\ref{eq:Pstarrhomexch}).  The degree of agreement indicates the
domain where the pseudoscalar meson exchange picture is effective.
The pion pole dominates the behavior at low momentum of the
pseudoscalar correlation.  For on-shell photons, the meson exchange
approximation is quite accurate for spacelike up to about \mbox{$K^2 =
1\;{\rm GeV}^2 \approx 50 m_\pi^2$}.  However with increasing $Q^2$,
one observes that the pion pole alone falls off with $K^2$ faster than
what is required to describe $A_{\rho P^*\gamma^*}(K^2,Q^2)$.  Similar
to the vector case, for spacelike photon momenta, the falloff of the
amplitude $A$ with $K^2$ is {\it slower} than that of a pseudoscalar
meson dominance model; the meson exchange picture thus has a very
limited domain of applicability.

In the bottom panel of Fig.~\ref{fig:ffinhomps} we display the
generalized form factor $G_{\rho\pi^*\gamma^*}(K^2,Q^2)$ as a function
of the photon momentum $Q^2$ for several values of the virtual
pseudoscalar $K^2$.  The physical form factor at the pseudoscalar
meson mass-shell, $F_{\gamma^*\rho\pi}(Q^2)$, is shown by the
continuous curve.  The rise at timelike photon momentum is due to the
vector pole in the photon-quark vertex.  This figure illustrates that
the form factor as a function of $Q^2$ becomes harder as the momentum
$K^2$ of the spacelike pseudoscalar correlation is increased.  For
spacelike $K^2 < 0.1\;{\rm GeV}^2\approx 5 m_\pi^2$ the meson exchange
approximation is quite good up to $Q^2 \approx 3~{\rm GeV}^2$, but
with the larger virtuality $K^2 > 0.8\simeq 40 m_\pi^2$ the error in
the meson dominance assumption has grown to almost a factor of $2$ at
$Q^2 \approx 3~{\rm GeV}^2$.  No standard phenomenological form factor
for the meson-nucleon coupling, $g_{PNN}(K^2)$, can compensate.  This
observation is evidently due to the fact that the $\bar{q}\gamma_5 q$
correlation does not continue to fall off with increasing spacelike
total momentum, but goes to a constant.

\section{\label{sec:summary}
Summary}
We have studied selected meson transition processes and associated
form factors within a model of QCD based on the Dyson--Schwinger
equations truncated to ladder-rainbow level.  The infrared structure
of the ladder-rainbow kernel is described by two parameters; the
ultraviolet behavior is fixed by the one-loop renormalization group
behavior of QCD.  Within the $u$ and $d$ quark sector we have obtained
the coupling constants for the radiative decays: \mbox{$\rho \to \pi
\gamma$}, \mbox{$\omega \to \pi \gamma$}, and \mbox{$\pi^0 \to \gamma
\gamma$}.  We have studied the form factors for the associated
transitions: \mbox{$\gamma^* \pi^0 \to \gamma$}, \mbox{$\gamma^* \pi^0
\to \gamma^*$}, \mbox{$\gamma^* \pi \to \rho$} and \mbox{$\gamma^*
\rho \to \pi$}.  The latter two processes are of interest as
contributors to meson electroproduction from hadronic targets away
from the $s$-channel resonance region.

We have exploited the fact that since a quark-gluon model can
dynamically produce the vector meson pole in the dressed photon-quark
vertex, the validity and effectiveness of using a meson exchange
picture in nearby momentum domains can be tested.  We find that for
the transition \mbox{$\gamma^* \pi^0 \to \gamma$}, the vector meson
resonance pole term extrapolated to the photon point produces an
estimate of the $\rho$ radiative decay coupling constant $g_{\rho \pi
\gamma}$ in terms of the $\pi$ decay coupling constant $g_{\pi \gamma
\gamma}$ that is accurate to within a few percent.  However at the
photon point and more generally for spacelike momentum, there is no
vector meson bound state; the object that occurs in such dynamics is a
vector $\bar{q}q$ correlation described by the dressed $\bar{q}
\gamma_\mu q$ vertex.  This particular meson-$\bar{q}q$ duality cannot
survive when the momentum of the vector object becomes sufficiently
large and spacelike because the dressed $\bar{q} \gamma_\mu q$ vertex
eventually becomes bare or perturbative, it does not fall off with
large spacelike momentum; that is, it cannot provide a fall-off like a
point meson propagator.  Nevertheless, for a large range of spacelike
momentum, we find the shape of the form factor for the transition
\mbox{$\gamma^* \pi^0 \to \gamma$} to be quite accurately described by
a monopole with mass scale $m_\rho$; that is also consistent with
analyses of the asymptotic behavior.  This is an example of the
empirical effectiveness of the simple VMD assumption being much
greater than its faithfulness to the underlying dynamics or physical
picture.

We have examined this issue further by considering form factors for
transition processes \mbox{$\gamma^* P \to \rho$} and \mbox{$\gamma^*
V \to \pi$} where $P$ and $V$ are virtual $\bar{q}q$ objects having
the quantum numbers of ground state pseudoscalar and vector mesons
respectively.  Such processes often arise in meson exchange models of
electroproduction of mesons from hadronic targets.  Our model of the
underlying quark-gluon dynamics is used to investigate the extent to
which the virtuality of $V$ and $P$ influences the corresponding form
factors that should be employed in meson exchange models, given that
the employed $t$-channel propagator is of the standard point meson
type.

Here we model the hadronic coupling to $V$ or $P$ as generated by a
point $\bar{q} \gamma_\mu q$ or $\bar{q} \gamma_5 q$ vertex
accompanied by a standard phenomenological meson-hadron form factor.
The physics of interaction and propagation of $V$ or $P$ towards the
$\gamma^* \pi$ or $\gamma^* \rho$ transition currents is implemented
by solution of the ladder BSE for the corresponding dressed versions
of the $\bar{q} \gamma_\mu q$ or $\bar{q} \gamma_5 q$ vertices.  We
contrast the results obtained this way with those from the meson
exchange picture in which the \mbox{$\gamma^* \pi \to \rho$} and
\mbox{$\gamma^* \rho \to \pi$} form factors are paired with the
corresponding point meson propagator.  In this way we obtain some
insight into the domain of applicability of the meson exchange
mechanism for these processes.

We find that near the photon point, the dependence of the amplitude
for \mbox{$\gamma^* V \to \pi$} upon the spacelike virtuality of $V$
is well described by the meson exchange picture out to at least
$5~{\rm GeV}^2$, which is the limit of our examination.  As we have
pointed out, there is indirect information that the agreement will
extend to asymptotic spacelike momenta.  However when both $\gamma^*$
and $V$ are more spacelike than about $1~{\rm GeV}^2$, the point meson
propagator for $V$ overestimates the falloff with virtuality of $V$ by
at least 50\% and the discrepancy increases with virtuality.  This
means that the \mbox{$\gamma^* V \pi$} form factor to be used in
effective point meson exchange models must have a dependence upon
virtuality of $V$ that compensates for the inadequacy of the point
meson propagator.  Similar conclusions are drawn for the
\mbox{$\gamma^* P \to \rho$} process.

Rather than model the coupling of the $\bar{q}q$ correlation to a
nucleon source by a $\bar{q}q$ dressed vertex seeded by a point vertex
and a phenomenological meson-nucleon form factor, a more exact
treatment would require a quark-gluon description of the nucleon
transition current.  It is a difficult task to combine such a
description with the meson transition form factors considered here.
Some progress in development of the required techniques within a DSE
approach can be found in Ref.~\cite{Bicudo:2001jq}.

\begin{acknowledgments}
We are grateful to T.S.H. Lee for a number of helpful discussions and 
questions that led to some of the present investigations. 
This work was funded by the National Science Foundation under Grant
Nos.\ PHY-9722429 and PHY-0071361 and by the Department of Energy
under Grant Nos.\ DE-FG02-96ER40947 and DE-FG02-97ER41048; it
benefitted from the resources of the National Energy Research
Scientific Computing Center.

\end{acknowledgments}

\bibliography{refsPM,refsPCT,refsCDR,refs}

\end{document}